\documentclass[final,5p,fleqn]{elsarticle}

\usepackage{graphicx}
\usepackage{graphicx}
\usepackage{hyperref}
\usepackage{amsmath}
\usepackage{url}
\usepackage{colordvi}

\journal{Physics Letter B} 

\begin{document}

\begin{frontmatter}

\title{On the Momentum Dependence of the Flavor Structure of the Nucleon Sea}

\author[a]{Jen-Chieh Peng}
\author[b]{Wen-Chen Chang}
\author[b]{Hai-Yang Cheng}
\author[b]{Tie-Jiun Hou}
\author[c]{Keh-Fei Liu}
\author[d,e]{Jian-Wei Qiu}

\address[a]{Department of Physics, University of Illinois at
Urbana-Champaign, Urbana, Illinois 61801, USA}

\address[b]{Institute of Physics, Academia Sinica, Taipei 11529, Taiwan}

\address[c]{Department of Physics and Astronomy, University of 
Kentucky, Lexington, Kentucky 40506, USA}

\address[d]{Physics Department, Brookhaven National Laboratory,
Upton, NY 11973, USA}

\address[e]{C.N. Yang Institute for Theoretical Physics and Department
  of Physics and Astronomy, Stony Brook University, Stony Brook, NY
  11794, USA}

\begin{abstract}
Difference between the $\bar u$ and $\bar d$ sea quark distributions
in the proton was first observed in the violation of the Gottfried sum
rule in deep-inelastic scattering (DIS) experiments. The parton
momentum fraction $x$ dependence of this difference has been measured
over the region $0.02 < x < 0.35$ from Drell-Yan and semi-inclusive
DIS experiments. The Drell-Yan data suggested a possible sign-change
for $\bar d(x)-\bar u(x)$ near $x \sim 0.3$, which has not yet been
explained by existing theoretical models. We present an independent
evidence for the $\bar d(x)-\bar u(x)$ sign-change at $x \sim 0.3$
from an analysis of the DIS data.  We further discuss the
$x$-dependence of $\bar d - \bar u$ in the context of meson cloud
model and the lattice QCD formulation.
\end{abstract}


\end{frontmatter}


It is now a well established fact that 
the $\bar u$ and $\bar d$ distributions in the proton are strikingly 
different. The first evidence for this difference came from the 
observation of the violation of the
Gottfried sum rule~\cite{gottfried} in a deep-inelastic scattering (DIS)
experiment by the NMC Collaboration~\cite{nmc}. The Gottfried sum rule,
$I_G\equiv \int^1_0 [F^p_2(x_B)-F^n_2(x_B)]/x_B~dx_B = 1/3$, is obtained
under the assumption of a symmetric $\bar u$ and $\bar d$ 
sea~\cite{gottfried}, where $x_B$ is the Bjorken variable and 
is effectively equal to parton momentum fraction $x$ 
probed in DIS using the leading order QCD factorization formalism 
of the structure function $F_2(x_B)$.  
The NMC measurement of $I_G = 0.235 \pm 0.026$
implies that this assumption is invalid with an $x$-integrated difference of 
$\int^1_0 [\bar d(x) - \bar u(x)] dx =0.148\pm 0.039$. 

The NMC result was subsequently
checked using two independent experimental techniques. From measurements
of the Drell-Yan cross section ratios of $[\sigma(p+d)]/[\sigma(p+p)]$, the
NA51~\cite{na51} and the Fermilab E866~\cite{e866} experiments
measured $\bar d/\bar u$ as a function of $x$ over the kinematic 
range of $0.015 < x < 0.35$. As shown in Fig. 1, the $\bar d / \bar u$ ratios
clearly differ from unity. 
From a semi-inclusive DIS measurement, the HERMES collaboration also
reported the observation~\cite{hermes98} of
$\bar d(x) - \bar u(x)\neq 0$, consistent with the
Drell-Yan results.

The $\bar d(x) / \bar u(x)$ data obtained from the Drell-Yan
experiments have provided stringent constraints for parametrizing the
parton distribution functions (PDFs).  Figure 1 compares the data
measured at $Q^2 = 54$ GeV$^2$ from Fermilab E866 with
parametrizations of several PDFs. The E866 data show the salient
feature that $\bar d / \bar u$ rises linearly with $x$ for $x < 0.15$
and then drops as $x$ further increases. At the largest value of
$x~(x=0.315)$, the $\bar d / \bar u$ ratio falls below unity, albeit
with large experimental uncertainty. This intriguing $x$-dependence of
$\bar d / \bar u$ is reflected in recent PDFs including
CTEQ6~\cite{cteq6}, CT10~\cite{ct10}, MSTW08~\cite{mstw08}, and
{JR14~\cite{jr14}}. However, for the CTEQ4M~\cite{cteq4m} PDF,
which predated the E866 data, the $\bar d / \bar u$ ratios at large
$x$ are not well described by the parametrizations.  In particular,
$\bar d(x)/ \bar u(x)$ remains greater than unity, or equivalently,
$\bar d(x) - \bar u(x) > 0$, at all $x$.  The parametrizations of the
more recent PDFs are sufficiently flexible to accommodate a
sign-change for $\bar d(x) - \bar u(x)$ at $x \sim 0.3$, as suggested
by the E866 data.
 
\begin{figure}[t]
\includegraphics[width=0.44\textwidth]{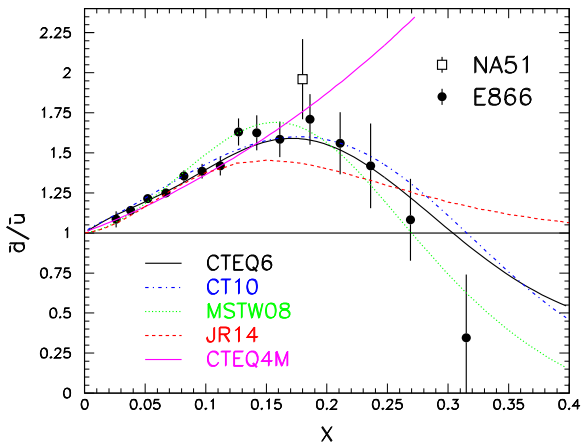}
\caption{Ratio of $\bar d(x)$ over $\bar u(x)$ versus Bjorken-$x$ 
from experiments NA51~\cite{na51} and E866~\cite{e866}.
Parametrizations from several parton distribution functions
are also shown.} 
\label{dbovub}
\end{figure}
Many theoretical models have been put forward to explain the surprisingly
large difference between $\bar d(x)$ and $\bar u(x)$. For reviews of various
theoretical models, see references~\cite{tony,kumano,vogt,garvey,peng_qiu}. 
While these models can explain the enhancement of $\bar d$
over $\bar u$ involving various mechanisms such as meson cloud,
chiral-quark, intrinsic sea, soliton, and
Pauli-blocking, none of them predicts that the $\bar d / \bar u$ ratio
falls below unity at any value of $x$~\cite{garvey}. 
In order to understand the origin of the sea-quark flavor structure, 
it is important to improve 
the accuracy and to extend the kinematic coverage of the $\bar d / \bar u$ 
measurement to the $x > 0.3$ region. This is the goal
of an ongoing Fermilab Drell-Yan experiment, E906~\cite{e906}, and
a proposed experiment~\cite{p04} at the J-PARC facility.
The $x$-dependence of $\bar d / \bar u$ (or the
related quantity $\bar d - \bar u$) at large
$x$ remains a topics of much interest both theoretically and
experimentally.

In this paper we address the intriguing 
possibility that $\bar d - \bar u$
changes sign at the $x \sim 0.3$ region. We first show that an independent
experimental evidence for this sign-change, other than the one shown in Fig. 1
from the Drell-Yan data, comes from an analysis of the 
NMC DIS data. 
We then discuss the significance of this sign-change and the stringent
constraint it imposes on theoretical models. 
We also discuss the implications on the $x$-dependence of
$\bar d - \bar u$ using 
the lattice QCD formulation for the sea-quark parton distributions.
Future measurements of $\bar d(x) / \bar u(x)$ at
$x > 0.25$ in Drell-Yan experiments could provide
strong constraints and new insights on the origins of the
flavor structure of the proton's sea.

\begin{figure}[t]
\includegraphics[width=0.44\textwidth]{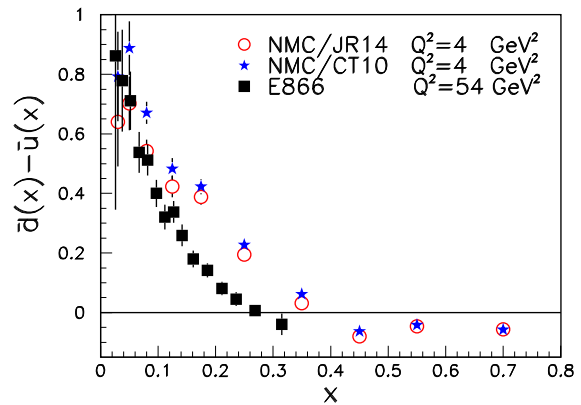}
\caption{Values of $\bar d(x) - \bar u(x)$ at $Q^2$ = 4 GeV$^2$ 
evaluated using Eq.~(\ref{eq:f2pn2}), as discussed in the text. 
The open circles and filled stars correspond to results obtained
with the JR14 and CT10 PDFs, respectively. Also shown 
are the values of $\bar d(x) - \bar u(x)$ 
at $Q^2$ = 54 GeV$^2$ from the Fermilab E866 experiment.}
\label{dbmud}
\end{figure}

The NMC measurement of the Gottfried
sum involves the $F_2$ structure functions on proton
and neutron. In terms of QCD factorization, we have at the leading order 
in $\alpha_s$,
\begin{equation}
F^p_2(x) - F^n_2(x)= \frac{1}{3} x[u(x) + \bar u(x) - d(x) - \bar d(x)],
\label{eq:f2pn1}
\end{equation}
\noindent where $x=x_B$ was used at this order.
Equation~(\ref{eq:f2pn1}) is obtained under the usual assumption
of charge symmetry of parton distributions and the equality of 
heavy-quark ($s, c, b$) distributions in proton and neutron. Note that
the $Q^2$ dependence in $F^{p,n}_2(x,Q^2)$ and parton 
distributions $q(x,Q^2)$ is implicit. {The magnitude of order $\alpha_s^1$ and $\alpha_s^2$ perturbative QCD effect is estimated to be small, on the order of 0.2\% at $Q=10$ GeV~\cite{kataev83}.}
From Eq.~(\ref{eq:f2pn1}) and the definition of valence
quarks, $u_v(x)=u(x)-\bar u(x)$ and 
$d_v(x)=d(x)-\bar d(x)$, one readily obtains
the following expression:
\begin{equation}
\bar d(x) - \bar u(x) = \frac{1}{2} [u_v(x) - d_v(x)] 
- \frac{3}{2x}[F^p_2(x)-F^n_2(x)].
\label{eq:f2pn2}
\end{equation}
\noindent Equation~(\ref{eq:f2pn2}) shows that the $x$ dependence of
$\bar d - \bar u$ can be extracted from the NMC measurement
of $F^p_2(x)-F^n_2(x)$ and the parametrization of $u_v(x) - d_v(x)$
from various PDFs. To illustrate this, we show in Fig.~\ref{dbmud}
the values of $\bar d(x) - \bar u(x)$ at $Q^2$ = 4 GeV$^2$ using
Eq.~(\ref{eq:f2pn2}), where the first term of the right-hand side,
$u_v(x) - d_v(x)$, is taken  
from the NNLO JR14 parametrization~\cite{jr14} and the second term, 
$F^p_2(x) - F^n_2(x)$, is taken from 
the NMC data~\cite{nmc} at $Q^2$ = 4 GeV$^2$. The JR14 is a recent PDF where the nuclear corrections from the CJ group~\cite{CJ11} is implemented and $\bar d(x) - \bar u(x) > 0$ is assumed at all $x$ in the global analysis. We also show
in Fig.~\ref{dbmud} the values of $\bar d(x) - \bar u(x)$
at $Q^2 = 54$ GeV$^2$ (filled squares) derived by the
E866 Collaboration~\cite{e866}. The sign-change of $\bar{d}-\bar{u}$ 
at $x \sim 0.3$ as indicated by the E866 data is clearly consistent 
with the behavior of open circles obtained by using Eq. (2) based 
on the NMC data and the JR14 PDFs.
Although the JR14 uses a parametrization of $\bar{d}-\bar{u}$ 
that is positive at all $x$, as shown in Fig. 1, we demonstrated 
in Fig. 2 that NMC data together with the valence quark 
distributions of JR14 could lead to a sign-change of 
$\bar{d}(x)-\bar{u}(x)$ distribution at $x\sim 0.3$.
We have also performed calculations with other sets of recent PDFs, 
obtained very similar results and reached the same conclusion.  
In Fig. 2, we show, for example, the values of 
$\bar{d}(x)-\bar{u}(x)$ (filled stars) obtained by using 
$u_v(x)-d_v(x)$ of the CT10 PDF parametrization~\cite{ct10} along 
with the same NMC data.  The values of 
$\bar{d}(x)-\bar{u}(x)$ obtained by using CT10 and 
JR14 are practically identical for $x > 0.2$.
This finding is effectively a consequence of the fact that
the $u_v(x) - d_v(x)$
distribution in Eq.~(\ref{eq:f2pn2}) is well constrained by
QCD global fit of the extensive DIS and hadronic scattering data.

\begin{figure}[t]
\includegraphics[width=0.5\textwidth]{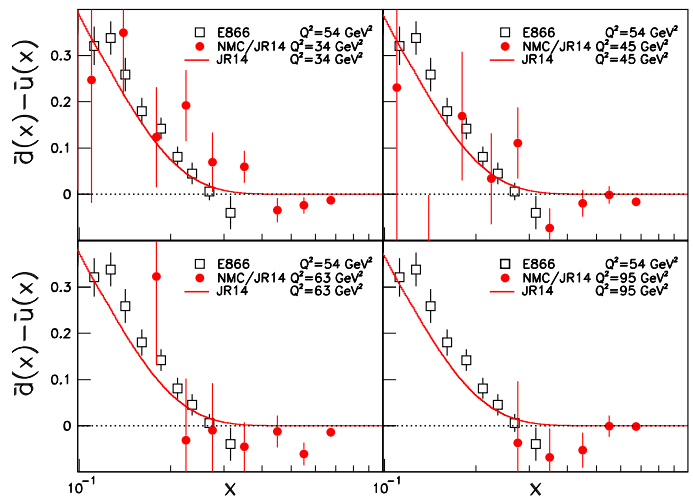}
\caption{Values of $\bar d(x) - \bar u(x)$ evaluated using
Eq.~(\ref{eq:f2pn2}) and the NMC data~\cite{nmc97,nmc95} 
of $R(x)$ and $F^d_2(x)$ at the four largest values 
of $Q^2$. {The JR14 parametrization for $u_v(x)-d_v(x)$ at the corresponding $Q^2$ and the ratio of deuteron to isoscalar nucleon structure functions $r_N^d(x)$ of the CJ12mid set at $Q^2 = 100$ GeV$^2$~\cite{CJ12} are used.}
The values of $\bar d(x) - \bar u(x)$
from E866 measurement at $Q^2 = 54$ GeV$^2$ are also shown. The
solid curves are $\bar d(x) - \bar u(x)$ from JR14.}
\label{dbmud_new}
\end{figure}

Although Fig.~\ref{dbmud} shows similar trends for the $x$-dependence of 
$\bar d - \bar u$ extracted from the
E866 Drell-Yan and the NMC DIS data, these two data
sets correspond to two different $Q^2$ scales. A more direct
comparison can be obtained by analyzing the final results published
by the NMC collaboration on the ratio 
$R(x) = F_2^d(x)/F_2^p(x)$~\cite{nmc97}. 
{The values of $F_2^p(x)-F_2^n(x)$ could be calculated from
$2F_2^d(x) * (1/R(x)-1/r_N^d(x))$, by using the parametrization of 
$F_2^d(x)$ of Ref.~\cite{nmc95}. The $r_N^d(x)$ is the ratio of deuteron to isoscalar nucleon structure functions $F_2^d(x) = r_N^d(x)*
(F_2^p(x)+F_2^n(x))/2$ and we use $r_N^d(x)$ of the CJ12mid set at $Q^2 = 100$ GeV$^2$~\cite{CJ12} for the evaluation.} Refs.~\cite{nmc97} and~\cite{nmc95} included not only 
additional data that were not available for NMC's earlier 
evaluation of the Gottfried sum~\cite{nmc}, but also the 
values of $R(x)$ at different bins of $Q^2$ ranging from 0.16 
to 99.03 GeV$^2$. The high $Q^2$ data makes it possible to 
compare the E866 Drell-Yan data on $\bar{d}(x) - \bar{u}(x)$ 
at $Q^2=54$~GeV$^2$ with that evaluated using Eq. (2) and 
NMC data at a similar $Q^2$.
As the E866 Drell-Yan data on 
$\bar d(x) - \bar u(x)$ correspond to $Q^2 = 54$ GeV$^2$, a 
comparison could be made by using the NMC data at similar $Q^2$.
The mean values of $Q^2$ for the four highest $Q^2$ bins
of NMC data are around 34, 45, 63, and 95 GeV$^2$.
Figure~\ref{dbmud_new} shows $\bar d(x) - \bar u(x)$ 
for these four values of $Q^2$ using
Eq.~\ref{eq:f2pn2} with the {JR14} parametrization of 
the valence quark distributions
and the NMC data~\cite{nmc97} for $F_2^p(x) -F_2^n(x)$.
The uncertainties of both $R(x)$ and the parametrization of $F^d_2$ 
have been included in the evaluation of $F_2^p(x) -F_2^n(x)$. 
Figure~\ref{dbmud_new} shows that the values of 
$\bar d(x) - \bar u(x)$ at $x>0.3$ are mostly negative with
the mean values of $-0.009 \pm 0.006, -0.012 \pm 0.006,
-0.016 \pm 0.008,$ and $-0.001 \pm 0.008$, respectively, for the four
$Q^2$ bins.
The agreement between the E866 and NMC results is
now improved when compared with Fig.~\ref{dbmud}. In particular, 
both the NMC and the
E866 experiments show evidence that $\bar d(x) - \bar u(x)$ changes
sign at $x\sim 0.3$.

Since both the NMC data and the E866/NA51 Drell-Yan data are included
in recent global fits for determining the parton distributions, it is
conceivable that the NMC data have already played a role
in constraining the behavior of $\bar d(x) - \bar u(x)$ at large $x$.
Nevertheless, the possible sign-change of $\bar d(x) - \bar u(x)$
for $x\sim 0.3$ has only been attributed in the literature
to the E866 data, which
have large uncertainty at the highest $x$ region. We show that an
independent indication for this sign-change is already provided
by the NMC DIS data, which were obtained prior to the E866 Drell-Yan
data. 

The significance of the sign-change of $\bar d(x) - \bar u(x)$ for
$x > 0.3$, if confirmed by future experiments, is that
it would severely challenge existing theoretical models which
can successfully explain $\bar d(x) - \bar u(x)$ 
at $x<0.25$, but predict no 
sign-change at higher $x$. Take for example the 
meson-cloud model~\cite{thomas83,henley90,kumano91,hwang91},
which treats proton as
a linear combination of a bare proton plus pion-nucleon and
pion-delta Fock states:

\begin{eqnarray}
|p\rangle & \to & \sqrt{1-a^2-b^2}~|p_0\rangle \nonumber \\
& & + a~\Large{[}-\sqrt{\frac{1}{3}}
|p_0\pi^0\rangle + \sqrt{\frac{2}{3}} |n_0 \pi^+\rangle\Large{]} \nonumber \\
& & +b~\Large{[}\sqrt{\frac{1}{2}}|\Delta^{++}_0 
\pi^-\rangle - \sqrt{\frac{1}{3}}
|\Delta^+_0\pi^0\rangle + \sqrt{\frac{1}{6}}|\Delta^0_0\pi^+\rangle\Large{]}.  \nonumber \\
& & \mbox{                                          }
\label{eq:meson}
\end{eqnarray}

\noindent The subscript zeros denote bare baryons with flavor symmetric
seas. The $\bar u$ and $\bar d$ seas have contributions from the
valence antiquarks of the pion cloud, i.e., $\bar d$ in
$\pi^+$ and $\bar u$ in $\pi^-$. The pion-nucleon amplitude 
is larger than the pion-delta amplitude ($a>b$) due to the
heavier mass for the $\Delta$. The excess of $\bar d$
over $\bar u$ arises because of the dominance of the
$n_0 \pi^+$ configuration over the less probable $\Delta_0^{++}\pi^-$
configuration. This leads to an overall excess of $\bar d$ over
$\bar u$. Moreover, the $x$ distribution for $\bar u$ is softer than
that of $\bar d$, since $\pi^-$ in the $\Delta_0^{++} \pi^-$ configuration
carries a smaller fraction of the proton's momentum
than $\pi^+$ in the $n_0 \pi^+$
configuration. As a consequence, $\bar d(x) - \bar u(x)$ remains
positive and does not change sign at large $x$. The same conclusion
can be obtained for the chiral 
quark model~\cite{Eichten92,cheng95,szczurek96}, in which the pions
couple directly to the constituent quarks. Since there are two
$u$ quarks coupling to $\pi^+$ ($u \to \pi^+ + d$) and only
one $d$ quark coupling to $\pi^-$ ($d \to \pi^- + u$),
the larger probability for the $\pi^+$ meson cloud relative to the
$\pi^-$ cloud would lead to $\bar d > \bar u$ for all $x$.
Similar conclusions can be obtained in the intrinsic sea 
model~\cite{chang11}, the chiral-quark soliton model~\cite{pobylitsa99},
and the statistical model~\cite{soffer}.

\begin{figure}[tbp]
\centering
\includegraphics[width=0.18\textwidth]{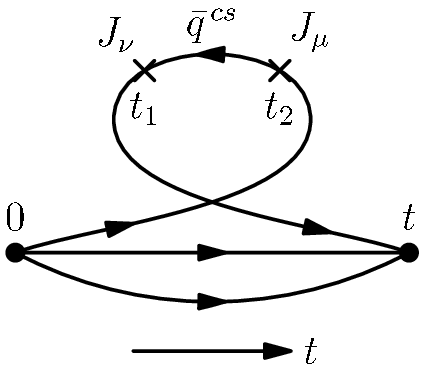}
\hskip 0.5cm
\includegraphics[width=0.18\textwidth]{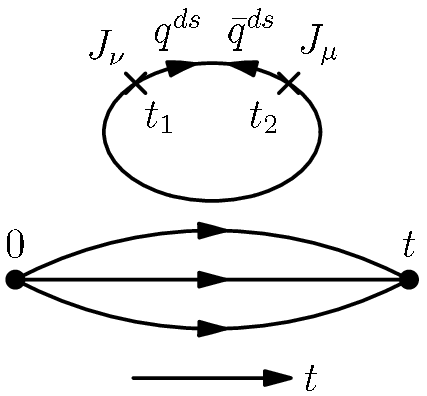}
\caption{Two gauge invariant and topologically distinct diagrams
for (a) connected sea (left graph) and (b) disconnected sea (right graph).}
\label{fig_liu_pdf}
\end{figure}

To shed some light on the $x$-dependence of $\bar d - \bar u$, we 
consider the origins of sea quarks in the lattice QCD approach.
There are two sources for the $\bar d$ and $\bar u$ seas in the
path-integral formalism of the hadronic tensor defining the 
structure function $F_2(x)$~\cite{liu00}, as
shown in the two gauge-invariant and topologically distinct
diagrams in Fig.~\ref{fig_liu_pdf}. One is the connected sea (CS) 
from the connected insertion diagram (Fig.~\ref{fig_liu_pdf}(a)) and the other
is the disconnected sea (DS) from the disconnected insertion 
diagram (Fig.~\ref{fig_liu_pdf}(b)). For the 
case with isospin
symmetry, i.e. $m_u = m_d$, it is shown~\cite{ld94} that the DS does
not distinguish $\bar{u}$ from $\bar{d} $. Hence, the 
$\bar{u}(x), \bar{d}(x)$
difference must originate solely from the CS (Fig.~\ref{fig_liu_pdf}(a)).
It is well known that sea quark distribution generated by
the disconnected diagram is a steeply falling
function of momentum fraction $x$, because the gluon radiated from the
initial quark line is dominantly soft due to the $\propto 1/x$ behavior
of the splitting kernel. In contrast, sea quarks generated by the connected 
diagram have an $x^{-1/2}$ behavior at small $x$ and are 
most relevant in the medium and large $x$ region.
While lattice QCD so far could only generate the moments rather than the
$x$-dependence of quark distributions, a first attempt to separate the
CS and DS components of the $\bar u(x) + \bar d(x)$ was reported 
recently~\cite{liu12}. The extracted CS and DS for
$\bar u(x) + \bar d(x)$~\cite{liu12}
have a distinct $x$ dependence in qualitative agreement with
expectation. The DS dominates the small $x$ region (i.e. $x < 0.05$)
while the CS dominates the $x > 0.05$ region.

It is instructive to consider $\bar{d}(x) - \bar{u}(x)$ in three
different $x$ regions. At small $x$, the DS with small $x$ behavior of
$x^{-1}$ dominates. For $Q^2 = 2.5~{\rm GeV^2}$ where
the CS and DS are explicitly separated~\cite{liu12}, this is the
region where $x < 0.05$. Since the only difference between
$\bar{u}^{DS}$ and $\bar{d}^{DS}$ is the $u/d$ mass difference
which is much smaller than the scale for the validity of the
parton picture, we expect the difference between them due to
isospin symmetry breaking to be very small. In the mid-$x$ region
(from $x = 0.05$ to $x \sim 1/3$),
dominated by the CS with a $x$-dependence of 
$x^{-1/2}$, the Fock space wavefunction of the quarks in the nucleon
is important. It is in this region that the DIS
and Drell-Yan experiments reveal that $\bar{u}(x) < \bar{d}(x)$.
The dominance of the CS at this region of $x$ suggests
a greater chance for the
CS partons to share the momentum with the valence quarks resulting in the
meson-baryon configurations. Hence the pronounced feature of
$\bar{u}(x) < \bar{d}(x)$ in this $x$ region can be understood in
terms of the pion cloud model.

\begin{figure}[tbp]
\centering
\includegraphics[width=0.3\textwidth]{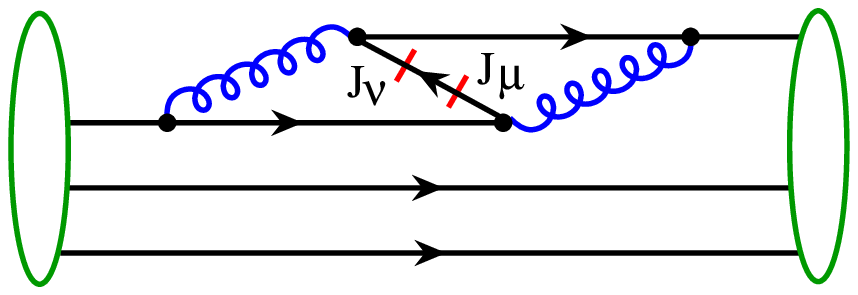}
\vskip 0.05in
time \ $\longrightarrow$
\vskip 0.05in
\includegraphics[width=0.3\textwidth]{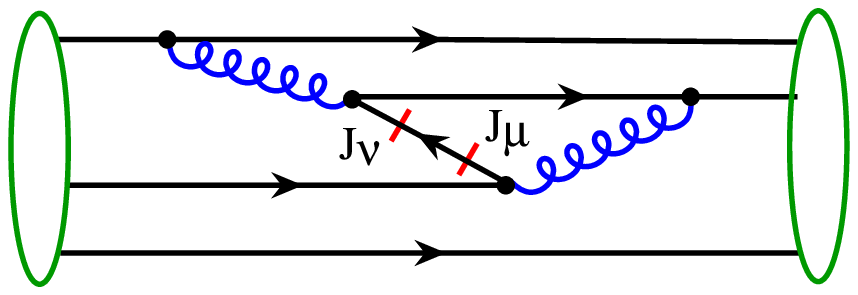}
\caption{QCD quantum fluctuation capable of generating
connected $\bar u(x)$ or $\bar d(x)$, 
involving one (top) or two (bottom) valence quarks, 
which could lead to more $\bar u(x)$ than $\bar d(x)$.}
\label{fig:ubar2dbar}
\end{figure}

At even larger $x$ ($x > 1/3$) the nature of the sea quarks is expected to be
strongly influenced by the valence quarks. Intuitively, the
connected sea diagram provides a natural mechanism for generating
more $\bar u(x)$ than $\bar d(x)$ at this region, since there are
two $u$ valence quarks capable of generating $\bar u$ quarks 
as shown in Fig.~\ref{fig_liu_pdf}. Figure 5 shows specific examples
of diagrams responsible for generating antiquarks from one (top) 
or two (bottom) valence quarks. 
In Fig.~\ref{fig:ubar2dbar}, the antiquark mode of the QCD quantum 
fluctuation of a quark is probed by the currents $J_\mu$ and $J_\nu$.
The quantum fluctuation could be thought as a time sequence of four steps:
1) fluctuation of a valence quark into a quark and a highly virtual gluon, 
2) a quick splitting of the gluon into a quark and antiquark pair, 
3) annihilation or recombination of the quark and the newly produced antiquark 
into a highly virtual gluon, which is then, 4) absorbed by the quark.
Since both valence $u$ and $d$ quarks can go through the same QCD
quantum fluctuation to generate $\bar u$ and $\bar d$ quarks, the mechanism
depicted in Fig.~\ref{fig:ubar2dbar} could generate about twice of $\bar u(x)$ 
over $\bar d(x)$ due to the 2-to-1 ratio of valence quarks.
But, this fluctuation is the most probable only if partons involved have 
an excellent coalescence, and therefore, it should be very short-lived.  
That is, it is unlikely to generate enough imbalance between $\bar{u}$ and 
$\bar{d}$ to compete with what could be generated by the 
pion cloud or other mechanisms/models at small-$x$.  
However, this mechanism is not very sensitive to the 
parent quark's momentum fraction $x$, and would become 
relevant when the imbalance generated by other mechanisms/models 
dies away at large $x$. {It is noted that $\bar u > \bar d$ was also suggested by a model calculation examining the antisymmetrization effect of the nucleon sea arising from gluon exchange between confined valance quarks~\cite{steffens97}.} The data indicates that such transition takes place at $x\sim 1/3$.
A detailed calculation of $\bar u(x)$ and $\bar d(x)$ in terms of connected
(or recombination) diagrams, like that in Ref.~\cite{Close:1989ca}, 
is beyond the scope of this letter, and will be presented later.

In summary, we have discussed the importance of the possible
sign-change for $\bar d(x) - \bar u(x)$ at $x \sim 0.3$ for
understanding the flavor structure of the nucleon sea.  We present an
independent evidence for the $\bar d(x)-\bar u(x)$ sign-change at
large $x$ from an analysis of existing DIS data.  This sign-change
cannot be explained by any existing theoretical model on the nucleon
sea. Nevertheless, a qualitative explanation for the sign-change at
large $x$ is provided in the context of lattice QCD formalism. {Up
  to now, only the connection between the NMC data and the integral of
  $\bar d(x) - \bar u(x)$ has been discussed. The current work hopefully
  would lead to some dedicated studies by the various PDF groups to
  assess the impact of the NMC data on the $x$-dependence of $\bar d -
  \bar u$.}  We note that $\bar{d}(x) - \bar{u}(x)$ can be calculated on the lattice~\cite{liu00} from the
structure function of the hadronic tensor or the recently proposed
direct calculation via a Lorentz boost~{\cite{Ji,lin14}.  When
reliable results are obtained, they would provide a direct check on
the possible sign-change for $\bar d(x) - \bar u(x)$ at $x \sim
0.3$. New experimental information on the $x$-dependence of the $\bar
d - \bar u$ at large $x$, anticipated for future Drell-Yan
experiments, together with comprehensive global analyses would
be critical for understanding the origins of the flavor structure of
the nucleon sea.

\section*{Acknowledgments}

We acknowledge helpful discussion with Jiunn-Wei Chen and
Chien-Peng Yuan. This work was supported in part by the
National Science Council of the Republic of China and the
U.S. Department of Energy and National Science Foundation.

\end{document}